\documentclass[11pt,a4paper]{article}

\usepackage[comma]{natbib}
\usepackage{anysize}
\usepackage{color}
\usepackage{graphicx}

\usepackage{amsmath,amssymb,amsthm,amsopn,bm}

\theoremstyle{remark}

\renewcommand{\today}{\begingroup
\number \day\space  \ifcase \month \or January\or February\or
March\or April\or May\or June\or July\or August\or September\or
October\or November\or December\fi \space  \number \year \endgroup}

\theoremstyle{plain}

\newtheorem{teor*}{Teorema}

\theoremstyle{definition}

\marginsize{3cm}{3cm}{3cm}{3cm}

\title{The Modal Age of Statistics}

\author{Jos\'e E. Chac\'on\footnote{Departamento de
Matem\'aticas, Universidad de Extremadura, E-06006 Badajoz, Spain. E-mail:
{\tt jechacon@unex.es}}}

\begin{document}

\maketitle

\begin{abstract}
\noindent Recently, a number of statistical problems have found an unexpected solution by inspecting them through a ``modal point of view''. These include classical tasks such as clustering or regression. This has led to a renewed interest in estimation and inference for the mode. This paper offers an extensive survey of the traditional approaches to mode estimation and explores the consequences of applying this modern modal methodology to other, seemingly unrelated, fields.
\end{abstract}

\medskip
\noindent {\it Keywords:} mode estimation, mode testing, modal clustering, modal regression

\newpage

\section{Introduction}
\label{sec:1}

The mean-median-mode trio involves the three most frequently used measures of central tendency of a dataset. They are taught within the very first classes of any course on basic Statistics. However, they do not share the same degree of importance: the sample mean (or average) is normally well understood and employed in everyday situations, the sample median is also useful and easy to visualize, but the mode, usually defined as the value of the dataset having the highest frequency of appearance, looks like a more bizarre measure of location. This uneven treatment was already noted by \cite{dalenius1965}, but it keeps being present as of today, to some extent.

Indeed, when the dataset consists of realizations from a continuous random variable then all the observed values are different with probability one and, therefore, the mode does not even make much sense. This is so because such a definition of mode is well suited for discrete or even categorical random variables, but not for continuous data. For the latter, it is necessary to resort to the population level to give a reasonable concept of mode, which is defined as the value that maximizes the probability density function. The problem, of course, is that this definition of the population mode does not allow for a simple sample version, a fact that still makes it less accessible than the mean and median for the common audience.

\begin{figure}\centering
\includegraphics[width=.5\textwidth]{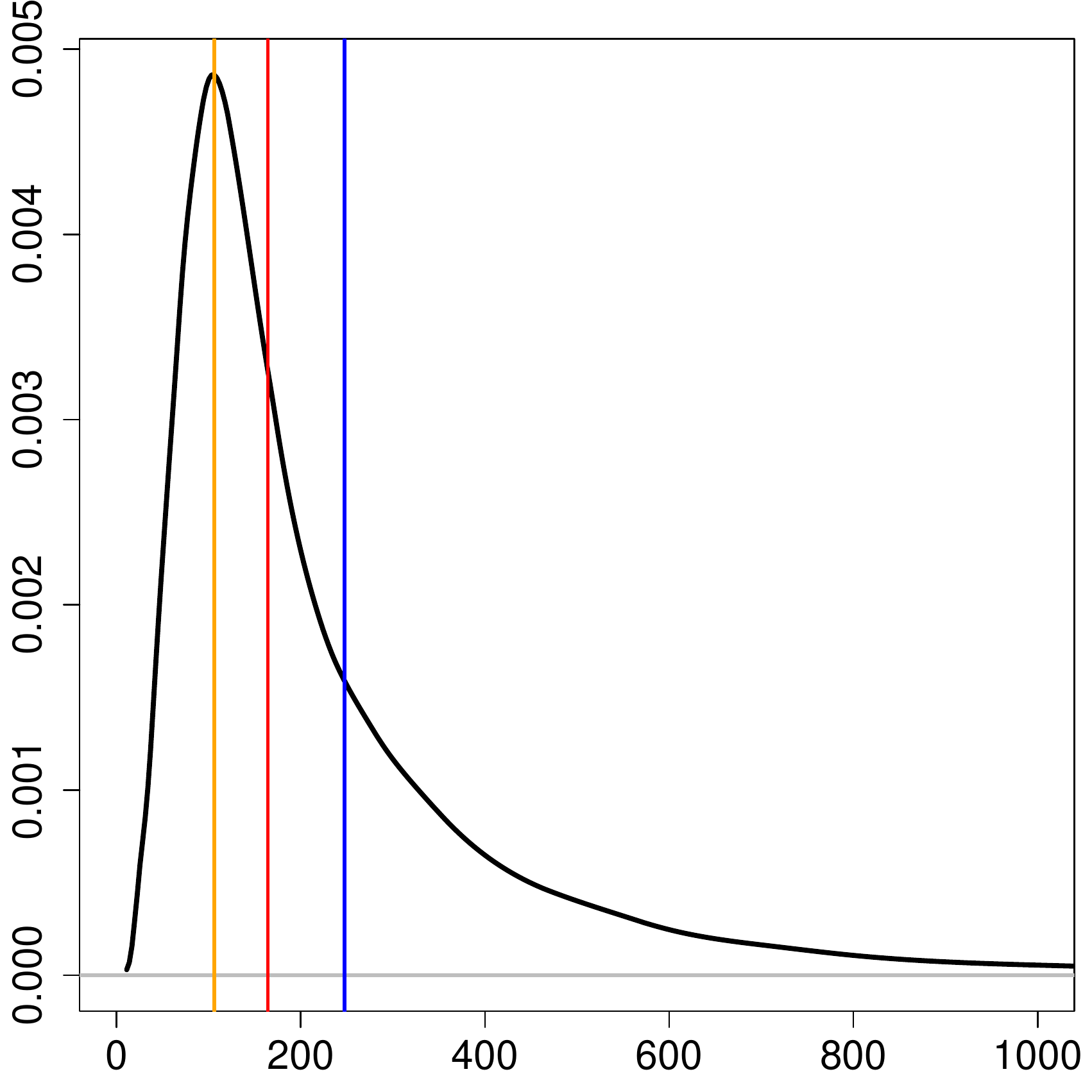}
\caption{Density estimate for the Phillipines total family income, in thousands of pesos. The blue, red and orange vertical lines show the (estimated) locations of the mean, median and mode, respectively.}
\label{fig:income}
\end{figure}

Nevertheless, it is not hard to find an easy-to-understand example where the mode appears as a more reasonable measure of location than the mean or the median. For instance, consider the random variable that records the total family income (TFI) per year, in pesos, in the Republic of Philippines in 2015. A dataset including $n=41544$ families is freely available from the Philippine Statistics Authority website\footnote{\tt http://openstat.psa.gov.ph/dataset/income-and-expenditure}. Figure \ref{fig:income} shows the (estimated) density of the TFI data. While the mean TFI is 247555.6 pesos, this surely appears as an unrealistic summary statistic for most Philippine families, since in fact more than 65\% of these TFI data are lower than 225000 pesos. In this case, the density mode is located at 106152 pesos, which probably describes more genuinely the situation of a typical family since, for instance, the shortest interval containing 50\% of the TFI data is $[57942, 182128]$. Here, the median TFI is 164079.5 pesos, so perhaps it still looks as a slightly too high summary value, given the previous remarks.

Obviously, the main reason for the failure of the mean and the median as measures of location in the previous example is the high skewness of the distribution of the TFI. Certainly, in a unimodal symmetric distribution the mean, mode and median all fall at the same point. So, in that case, it would not be necessary to distinguish between the three measures of location. However, symmetric distributions are not so often encountered in real data scenarios, and this fact has motivated a renovated and intense interest in the study of more flexible, possibly asymmetric distribution models in the last decades (see \citealp{genton2004}, and \citealp{azzalini2014}). For all these models, the mode emerges as a more natural measure of location than the mean or the median.

Thus, one of the goals of this paper is to provide, in an expository style, a brief review of the state of the art in mode estimation. This is the aim for Section 2, focusing on the simplest case where the mode is assumed to be the only global maximizer of the density. The scope of mode estimation can then be extended by considering not only the global mode, but also all the density local maxima that show up in a multimodal distribution. This generalization poses many additional statistical challenges, which are examined in Section 3. Moreover, recently the ``modal point of view'' has influenced not only those basic setups, but also apparently unrelated tasks like clustering or regression. So, in addition to the base cases, this paper also explores in Sections 4 and 5 the applications of this modern modal approach to other statistical problems. Some final conclusions and open questions are discussed in Section 6.

\section{The base case: estimation of a single mode}

The simplest setup that will be tackled here consists of estimating the mode of a density $f$ on $\mathbb R^d$ from an iid sample $\mathcal X=\{X_1,\dots,X_n\}$ with density $f$.

Any value $\theta\in\mathbb R^d$ such that $f(\theta)=\max_{x\in\mathbb R^d}f(x)$ is formally called a mode. This very definition, however, is not without problems and any quick-witted mathematician may readily pose problematic, and not too pathological, situations where such definition appears slippery (What is the mode of a uniform distribution? What if the density function is changed at a single point, while keeping the probability distribution unaltered?). In order to keep a lightweight reading as much as possible, 
it will be assumed that we are in a ``well-behaved situation'', meaning that density $f$ is continuous in a neighborhood of its unique mode $\theta$, as in \cite{abraham2003}. 

Following \cite{sager1978}, mode estimators can be classified as indirect and direct. Indirect estimators are defined as $\widehat\theta=\mathop{\rm argmax}_{x\in\mathbb R^d}\widehat f(x)$, where $\widehat f(x)$ is a suitable density estimator, whereas direct estimators are not based on estimating the density as a preliminary step. Ultimately, though, most direct estimators reveal an intrinsic connection with some kind of density estimator, but in the following we will review them separately.

\subsection{Direct estimators}

Direct estimators are surely less popular than their indirect counterparts, even if historically they were introduced at about the same time.

In the univariate case, a direct estimator is naturally defined as any point of an interval containing a high concentration of observations. Usually, the interval midpoint is employed. It is possible to proceed either by choosing the interval of a fixed length that contains the most observations, as in \cite{chernoff1964}, or by taking the shortest interval that contains a fixed proportion of the observations, as independently suggested by \cite{dalenius1965} and \cite{venter1967}. While these two estimators stand out as the two most prominent members of the class of direct estimators, it should be noted that Chernoff's estimator can be also viewed as an indirect estimator based on the naive kernel density estimator, and that \cite{moore1969} noted a connection between the Dalenius-Venter estimator and the nearest-neighbor density estimator as well.

To elaborate, Chernoff's estimator $\widehat \theta_{\rm C}$ can be defined as the midpoint of the interval of length $2a$ which contains a maximal number of observations. \cite{chernoff1964} showed that $\widehat \theta_{\rm C}\to\theta$ in probability if $a=a_n\to0$ and $na/\log n\to\infty$ as $n\to\infty$, provided
\begin{equation}\label{eq:well-separated}
\sup_{|x-\theta|>\delta}f(x)<f(\theta)\text{ for all }\delta>0.
\end{equation}
The assumptions on the interval length are used to ensure that the implicitly used naive kernel density estimator is uniformly consistent, and condition (\ref{eq:well-separated}) implies that the mode is  well separated, in the sense that the density cannot have far away spikes that approach $f(\theta)$ arbitrarily close. Hence, consistency follows as in the classical argmax theorem \citep[][Theorem 5.7]{vaart1998}. Heuristically, Chernoff also noted that optimally choosing $a$ of order $n^{-1/8}$ entails a rate of convergence of order $n^{-1/4}$, assuming that
\begin{equation}\label{condf}
\begin{split}
&\text{$f$ has a continuous third derivative in some neighborhood of $\theta$,}\\
&\text{with $f^{(2)}(\theta)<0$ and $f^{(3)}(\theta)\neq0$.}
\end{split}
\end{equation}
Asymptotic distribution theory for $\widehat \theta_{\rm C}$ is quite sophisticated, and has interest of its own \citep{groeneboom2001}. Chernoff formally showed that
\begin{equation*}
(na^2)^{1/3}(\widehat\theta_{\rm C}-\theta)\stackrel{d}{\to}\big\{2f(\theta)\big/f^{(2)}(\theta)^2\big\}^{1/3}T,
\end{equation*}
with $T=\mathop{\rm argmax}_{t\in\mathbb R}\{B(t)-t^2\}$, where $\{B(t)\}_{t\in\mathbb R}$ is a two-sided standard Brownian motion process with $B(0)=0$ and $\stackrel{d}{\to}$ denotes convergence in distribution.

On the other hand, for a given value of $k<n$, the Dalenius-Venter estimator $\widehat \theta_{\rm DV}$ is defined as the midpoint of the shortest interval containing $k$ of the sample observations. \cite{venter1967} showed that if $k=k_n$ is such that $k/n\to0$ as $n\to\infty$ and $\sum_{n=1}^\infty n\lambda^{k}<\infty$ for all $\lambda\in(0,1)$, then $\widehat \theta_{\rm DV}\to\theta$ almost surely, under mild conditions on $f$.
He also obtained the rate of convergence for $\widehat \theta_{\rm DV}-\theta$, which depends on the sharpness of the density around $\theta$, expressed in terms of the quantity $\alpha(\delta)$, which equals $f(\theta+\delta)/f(\theta+2\delta)$ for sufficiently regular densities. If there exist constants $\rho,r>0$ such that $\alpha(\delta)\geq1+\rho\delta^r$ for all small enough $\delta$, then by optimally taking $k=cn^{2r/(2r+1)}$ for some $c>0$ the convergence rate was shown to be of order $n^{-1/(2r+1)}(\log n)^{1/r}$. As noted by \citet[][p. 1451]{venter1967}, under the assumption (\ref{condf}) the condition on $\alpha(\delta)$ holds with $r=2$, and that suggests taking $k=cn^{4/5}$ to obtain a convergence rate of order $n^{-1/5}(\log n)^{1/2}$, which is slightly slower than for Chernoff's estimator. Venter also obtained the asymptotic distribution of $\widehat \theta_{\rm DV}$, which is related to Chernoff's distribution. If $k=cn^{4/5}$ then, under assumption (\ref{condf}),
$$(n^{-1}k^2)^{1/3}(\widehat\theta_{\rm DV}-\theta)\stackrel{d}{\to}f(\theta)\big\{2\big/f^{(2)}(\theta)^2\big\}^{1/3}T.$$

In a similar spirit, \cite{robertson1974} proposed an iterative scheme to estimate the mode: fix a proportion $p\in(0,1)$ and start by finding the smallest interval containing at least $np$ observations, then find the smallest subinterval within the former that contains at least $np^2$ observations; repeat this procedure until some stopping criterion is met, so that the final interval contains at least $np^\ell$ observations and select the midpoint $\widehat \theta_{\rm RC}$ of such final interval as the estimate of $\theta$. \cite{robertson1974} showed that $\widehat \theta_{\rm RC}$ is strongly consistent for $\theta$ if $p=p_n\to p_\infty$ for some $p_\infty\in(0,1)$ and $\ell=\ell_n\to\infty$ as $n\to\infty$, provided $f$ is unimodal (i.e., if $f(x)$ is increasing for $x\leq\theta$ and decreasing for $x\geq\theta$). More recently, \cite{bickel2006} analyzed $\widehat \theta_{\rm RC}$ from a robustness point of view and recommended using $p=1/2$ since it yields the choice with the highest breakdown point. They called this estimator half-sample mode, and showed that it compares favorably with many others direct estimators that are reviewed in their simulation study.

Still in the univariate case, \cite{grenander1965} proposed a completely different direct mode estimator, based on the $k$-spacings; that is, on the differences $D_{i,k}=X_{(i+k)}-X_{(i)}$, $i=1,\dots,n-k$, for some $k<n$, where $X_{(1)},\dots,X_{(n)}$ denote the order statistics. Given $k<n$ and $1<p<k$, Grenander's estimate is defined as $\widehat \theta_{\rm G}=\widehat B/\widehat A$, where $\widehat B=n^{-(p+1)}\sum_{i=1}^{n-k}D_{i,k}^{-p}(X_{(i+k)}+X_{(i)})/2$ and $\widehat A=n^{-(p+1)}\sum_{i=1}^{n-k}D_{i,k}^{-p}$; that is, $\widehat \theta_{\rm G}$ is a weighted average of the midpoints of intervals containing $k+1$ observations. For fixed $k$ and $p$, Grenander showed that $\widehat \theta_{\rm G}\to\theta_p$ in probability, where $\theta_p=\int_{-\infty}^\infty xf^{p+1}(x)dx\big/\int_{-\infty}^\infty f^{p+1}(x)dx$. The rationale that allows to consider $\widehat\theta_{\rm G}$ as a mode estimator is that $\theta_p\to\theta$ as $p\to\infty$, under appropriate regularity conditions. Although $\widehat\theta_{\rm G}$ seemed to work well in most simulation studies, its theoretical performance as mode estimator remained elusive until an impressive paper by \cite{hall1982} fully described its convergence rate and asymptotic distribution. Hall showed that if $p=p_n$ and $k=k_n$ are such that $p\to\infty$ and $k\to\infty$, with $p^2/k\to0$ and $p^4k^3/n^3+k^3/n^2\to0$, then by optimally taking $p$ of order $n^{2/7}$ the convergence rate of $\widehat \theta_{\rm G}-\theta$ is of order $n^{-2/7}$ when $f$ satisfies similar but slightly stronger conditions than (\ref{eq:well-separated}) and (\ref{condf}). This rate is faster than that of $\widehat\theta_{\rm C}$ and $\widehat\theta_{\rm DV}$. Moreover, he also obtained the asymptotic distribution, so that if $p=cn^{2/7}$ for some $c>0$, then
$$(np^{-3/2})^{1/2}(\widehat \theta_{\rm G}-\theta)\stackrel{d}{\to}v_{\rm G}Z+b_{\rm G},$$
where $Z$ is a standard normal random variable,  $v_{\rm G}^2=\{16\pi f(\theta)|f^{(2)}(\theta)|\}^{-1/2}$ and $b_{\rm G}=c^{-7/4}f(\theta)f^{(3)}(\theta)/\{2f^{(2)}(\theta)^2\}$.

Multivariate versions of these direct estimators are not easy to construct. The problem was first addressed by \cite{sager1978,sager1979}, who suggested a generalization of the Dalenius-Venter estimator defined as the central point of the minimum volume set, in a certain class, containing at least a certain proportion $p$ of the data. Such a class of sets intends to mimic the role of intervals in the univariate case to high dimensions. \cite{sager1979} initially considered the class of convex sets (this paper was written before, but published later), but for computational tractability reasons he recommended using hyper-rectangles or hyper-spheres in \cite{sager1978}. Recently, \cite{kirschstein2016} studied the finite sample behavior of Sager's multivariate mode estimator using convex sets and again, based on robustness considerations, recommended using $p=1/2$ as in \cite{bickel2006}.

\subsection{Indirect estimators}\label{sec:indest}

Any reasonable density estimator $\widehat f(x)$ yields an indirect mode estimator defined as $\widehat\theta=\mathop{\rm argmax}_{x\in\mathbb R^d}\widehat f(x)$. Indeed, the seminal paper by \cite{parzen1962} on kernel density estimation already addressed mode estimation as well.

The kernel density estimator is probably the most extensively studied density estimator. It is given by $\widehat f_{\rm K}(x)=n^{-1}\sum_{i=1}^nK_h(x-X_i)$, where $K$ is the kernel (an integrable function with unit integral), $h>0$ is the bandwidth or smoothing parameter and the scaled kernel is $K_h(x)=K(x/h)/h^d$. Recent monographs on kernel smoothing include \cite{tsybakov2009}, \cite{klemela2009}, \cite{horova2012} and \cite{chacon2018}.

Many authors have studied the kernel mode estimator $\widehat \theta_{\rm K}$, starting with the aforementioned work of \cite{parzen1962}, but one of the finest papers about its properties (in the univariate case) is surely \cite{romano1988}. There, it is shown that if $h=h_n\to0$ and $nh/\log n\to\infty$ as $n\to\infty$, then $\widehat \theta_{\rm K}$ is strongly consistent for $\theta$, provided $f$ is continuous in a neighborhood of $\theta$ and condition (\ref{eq:well-separated}) holds. Moreover, under assumption (\ref{condf}) Romano also showed that the mean-squared-error (MSE) optimal bandwidth is of order $n^{-1/7}$, which yields a convergence rate of order $n^{-2/7}$ for $\widehat \theta_{\rm K}-\theta$, the same as that of $\widehat \theta_{\rm G}$. In addition, if the bandwidth is taken to be $h=cn^{-1/7}$ for some $c>0$, then
$$(nh^3)^{1/2}(\widehat \theta_{\rm K}-\theta)\stackrel{d}{\to}v_{\rm K}Z-b_{\rm K},$$
where $v^2_{\rm K}=\{f(\theta)/f^{(2)}(\theta)^2\}R(K')$ and $b_{\rm K}=\frac12c^{7/2}\{f^{(3)}(\theta)/f^{(2)}(\theta)\}\mu_2(K)$, with $R(K')=\int_{-\infty}^\infty \{K'(x)\}^2dx$ and $\mu_2(K)=\int_{-\infty}^\infty x^2K(x)dx$.

As noted above, Chernoff's estimator $\widehat \theta_{\rm C}$ is the same as $\widehat \theta_K$ when the kernel is taken to be the density of the uniform distribution and $h=a$. However, the asymptotic theory for $\widehat\theta_{\rm C}$ is markedly different due to the fact that $R(K')=0$ for the uniform kernel. So mode estimation represents a noteworthy problem in kernel smoothing where the choice of the kernel indeed affects the convergence rate and asymptotic distribution of the resulting estimator, an unusual influence as compared to most scenarios.

Different convergence rates are found depending on the smoothness of the density $f$ and its flatness around the mode. Specifically, if $f$ is $\beta$-times continuously differentiable on a neighborhood of $\theta$, with $f^{(j)}(\theta)=0$ for $j=1,\dots,q-1$ and $f^{(q)}(\theta)<0$ for some even $q\leq\beta$, and $f^{(\beta)}(\theta)\neq0$, then \cite{vieu1996} showed that the convergence rate of $\widehat \theta_{\rm K}-\theta$ is of order $n^{-(\beta-1)/\{(q-1)(2\beta+1)\}}$. This means that convergence rates for kernel mode estimators are faster as $f$ is smoother and slower as $f$ is flatter around $\theta$. In fact, \cite{tsybakov1990} obtained the minimax convergence rate for mode estimation, showing that when $q=2$ no mode estimator (of any type) can have a convergence rate faster than $n^{-(\beta-1)/(2\beta+1)}$ under the above smoothness assumption on the density. So under condition (\ref{condf}), which implies that $q=2$ and $\beta=3$, the kernel mode estimator $\widehat \theta_{\rm K}$ is indeed rate-optimal.

The kernel mode estimator admits a natural generalization to the multivariate case, thanks to the existing developments of kernel density estimation for multivariate data. Its definition as the maximizer of the density estimate, however, involves a numerical minimization step which can be time-consuming in higher dimensions, so \cite{devroye1979} suggested a computationally simpler alternative defined as $\widetilde\theta_{\rm K}=\mathop{\rm argmax}_{x\in\mathcal X}\widehat f_{\rm K}(x)$, that is, searching among the observations $X_1,\dots,X_n$ for the one with the highest estimated density. \cite{abraham2003} studied this estimator, showing its consistency and convergence rates, and later they proved that its asymptotic distribution is the same as that of $\widehat\theta_{\rm K}$ \citep{abraham2004}, which was obtained in the multivariate case in \cite{mokkadem2003}.

Choosing the estimate among the data points is also quite convenient when the indirect estimator is based on the nearest-neighbor density estimate. Such a density estimate is defined as $\widehat f_{\rm NN}(x)=k/\{nv_d\|X_{(k)}(x)-x\|^d\}$, where $k$ is a natural number, $X_{(k)}(x)$ stands for the $k$-th nearest neighbor of $x$ among $X_1,\dots,X_n$, $\|\cdot\|$ represents the Euclidean norm and $v_d=\pi^{d/2}/\Gamma(1+d/2)$ denotes the volume of the unit ball in the Euclidean space $\mathbb R^d$ \citep{loftsgaarden1965}. Strong uniform consistency of $\widehat f_{\rm NN}$ was shown in \cite{devroye1977} whenever $k=k_n$ is such that $k/\log n\to\infty$ and $k/n\to0$, provided $f$ is uniformly continuous. So if in addition (\ref{eq:well-separated}) holds, that guarantees that $\widetilde\theta_{\rm NN}=\mathop{\rm argmax}_{x\in\mathcal X}\widehat f_{\rm NN}(x)$ is strongly consistent. More recently, \cite{dasgupta2014} studied the convergence rate of $\widetilde\theta_{\rm NN}-\theta$, and proved that it is of order $n^{-1/(d+4)}$ if $f$ is assumed to be locally quadratic around $\theta$.

On the other hand, in recent times there has been a growing interest in density estimation on shape-restricted density classes $\mathcal P$ \citep[see][]{groeneboom2014}. This is so because, for particular classes, it has been shown that it is possible to estimate the density via maximum likelihood (ML), whose main advantage relies on the fact that it does not require any tuning parameter to be chosen. So as before, if an ML density estimate $\widehat f_{\rm ML}$ is available, it is natural to consider its maximizer as an indirect mode estimator in that setting. Unfortunately, it can be shown that the ML estimator does not exist when $\mathcal P=\mathcal P_U$ is the class of all unimodal densities \citep{birge1997}. However, the ML density estimator exists within the subclass $\mathcal P_{LC}\subset\mathcal P_U$ of all log-concave densities (i.e., those densities whose logarithm is concave); see \cite{samworth2017} for a recent review on log-concave density estimation. Asymptotic theory for $\widehat \theta_{\rm ML}$, the indirect ML mode estimate within the class $\mathcal P_{LC}$, was developed in \cite{balabdaoui2009}, where it was shown that its convergence rate is of order $n^{-1/(2q+1)}$, with $q$ the order of the first nonvanishing derivative at $\theta$ (as above). Moreover, \cite{han2016} proved this rate is optimal for this problem.

\section{Many modes}

The methods in the previous section focus on the mode as the location of the global maximum of the density. Some methodologies even require the density to be unimodal, in the sense that, in addition, such a mode is the only local maximum of the density, as seems to be the case for the TFI data in Section \ref{sec:1}. However, it is common to use the denomination ``mode'' to refer to any local maximum of the density. Hence, any distribution whose density exhibits more than one local maximum is called multimodal.

\begin{figure}\centering
\includegraphics[width=.5\textwidth]{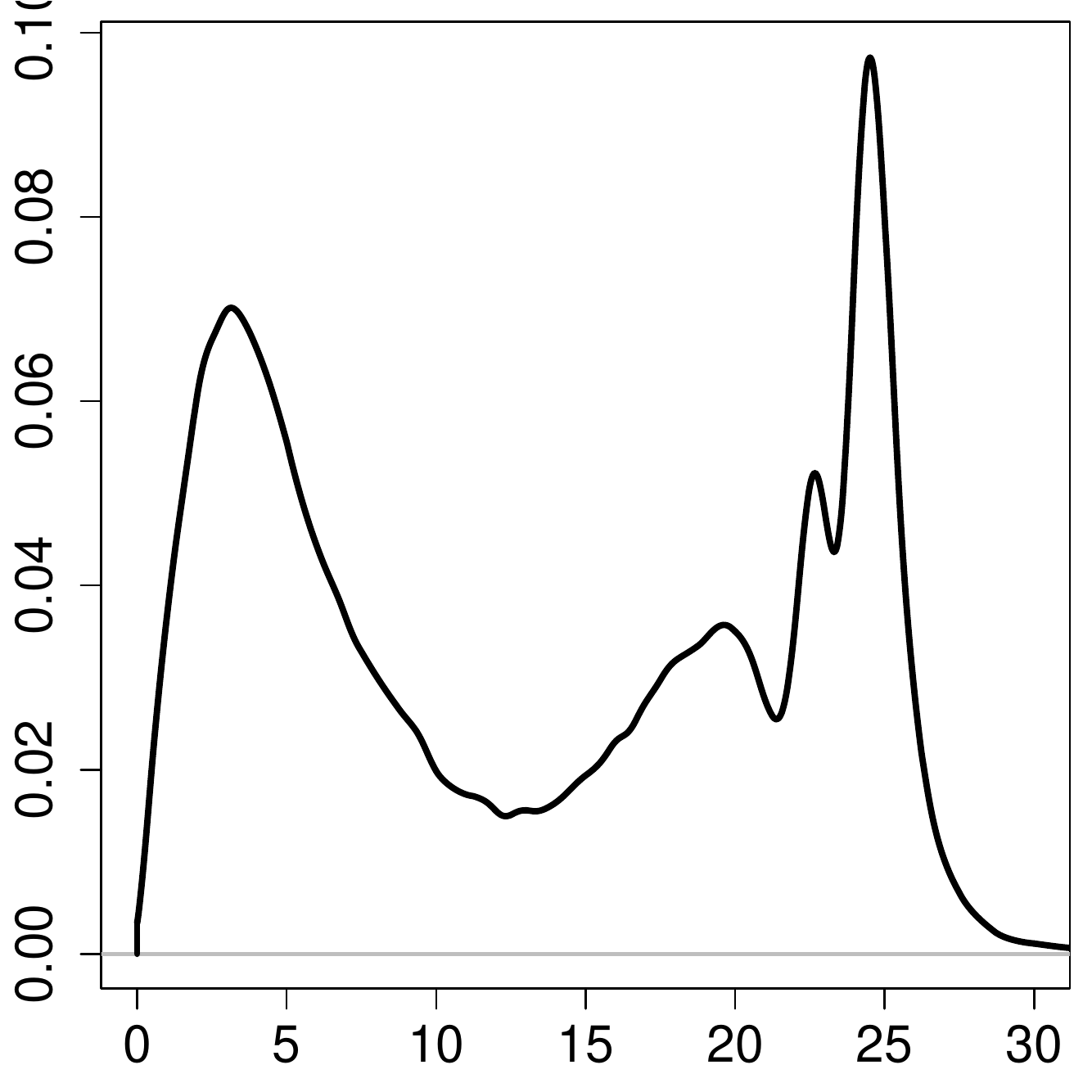}
\caption{Distribution of the distance to the basket of NBA shots during 2014-2015 season.}
\label{fig:NBA}
\end{figure}

Multimodal distributions further exemplify situations where the mean or the median may not seem to provide a sensible notion of centrality. For instance, Figure \ref{fig:NBA} shows a density estimate for the distance to the basket (in feet) of all shots made during the regular 2014-2015 season of the NBA. This a dataset comprising $n=128065$ shots, freely available from the Kaggle website\footnote{\tt https://www.kaggle.com/dansbecker/nba-shot-logs}. The density estimate presents four modes, which correspond to four different prominent shooting distances that the NBA players attempt. However, the sample mean and median, found at 13.57 and 13.70 feet respectively, are located just at a relatively unlikely shooting distance.

In this case, it could be argued that even the global mode, located at 24.51 feet, does not represent a central value of the distribution either, but it is rather the distance around which it is more likely to find a player shooting if somebody watches a basketball game at a random instant. {Not for nothing, the 3-point arc for NBA games is located at 23.75 feet from the basket.} In any case, the other three modes, located at 22.67, 19.62 and 3.15 feet respectively, are not far in terms of the chances of watching a shot from that distance (especially, the last one), which highlights the importance of performing a ``modal analysis'' of the distribution.

Multimodal distributions are usually understood to reflect the existence of several sub-populations within the distribution. This can be modeled through a mixture density, that is, via expressing the density as $f=\sum_{\ell=1}^L\pi_\ell f_\ell$. Here, the mixture weights $\pi_\ell>0$ are such that $\sum_{\ell=1}^L\pi_\ell=1$, and the component densities $f_\ell$ are employed to represent the different sub-populations \citep[see][]{mclachlan2000}. It should be noted, however, that the number of mixture components and the number of modes does not necessarily coincide, unless the components are sufficiently well separated, so that a modal point of view can be beneficial even for a mixture model approach. This will be further explored in Section \ref{sec:modclust} below.

Estimation of the density modes (i.e. the local maxima of the density) is usually accomplished by the same indirect approach as for a single mode; that is, by considering the local maxima of a density estimate $\widehat f$ as estimates of the true density modes. The characterization of local maxima in terms of the density derivatives links these mode estimators to the problem of density derivative estimation. Hence, more restrictive conditions are needed to ensure consistency and other asymptotic properties \citep{vieu1996}. However, despite apparently being a more difficult problem, it should be noted that minimax rates for the estimation of a single mode are, likewise, closely related to the achievable accuracy in density derivative estimation \citep{romano1988,tsybakov1990}.

\begin{figure}\centering
\includegraphics[width=.5\textwidth]{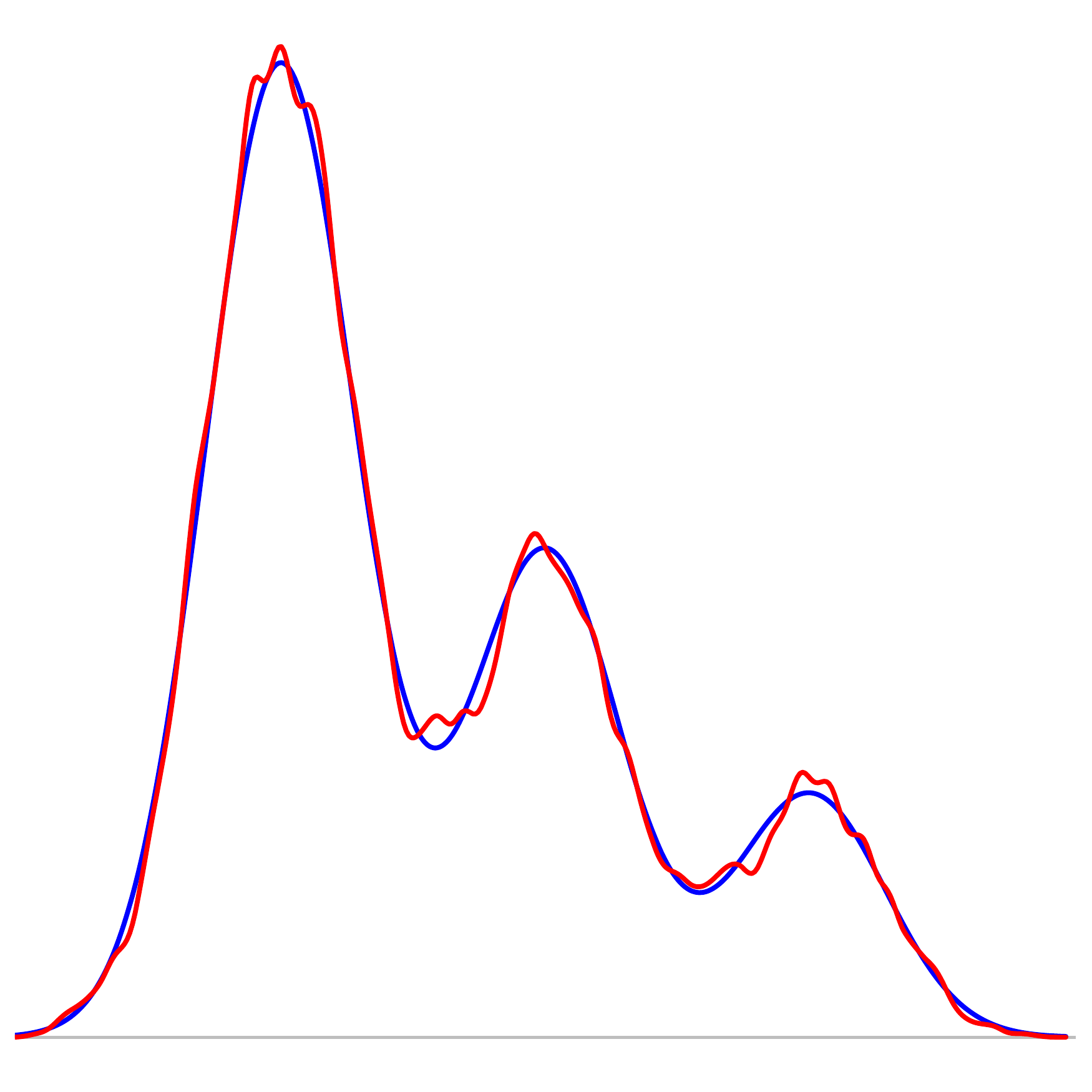}
\caption{Two close densities with a very different number of modes.}
\label{fig:2dens}
\end{figure}

A particularly difficult problem in this regard is the estimation of the true number of modes $\nu$. For instance, \cite{donoho1988} showed that only lower confidence bounds can be constructed for this problem, so that a procedure can be found that allows to conclude with 95\% confidence that the true distribution has {\it at least} three modes (say), but it is not possible to infer if the distribution has {\it at most} seven modes. Equivalently, it is possible to test the null hypothesis $H_0\colon\nu=n_0\in\mathbb N$ against the alternative $H_a\colon\nu>n_0$, but not against the reversed alternative. The intuition behind this negative result is the fact that, given any density, it is possible to find distributions with many more modes that are not too far from it, at least with respect to a certain ``testing distance'' that measures the chances of distinguishing between the two, based on a finite sample. For example, the two densities in Figure \ref{fig:2dens} are close, but while the blue curve shows three modes, the red one has eight. It does not seem possible, though, to find a close distribution with only two modes.

\cite{donoho1988} also provided an explicit construction of a lower confidence bound for $\nu$, while testing $H_0\colon\nu=n_0$ against $H_a\colon\nu>n_0$ was first considered by \cite{silverman1981}. A recent literature review about the testing problem can be found in \cite{ameijeiras2016}. The same authors also developed the R package {\tt multimode} \citep{ameijeiras2018}, in which most of the existing procedures were implemented for their practical use. From the estimation point of view, \cite{mammen1995} considered the number of modes of a kernel density estimator as an estimator of $\nu$, and derived its asymptotic behavior \citep[see][for the multivariate case]{konakov1997}.

In fact, the red curve in Figure \ref{fig:2dens} is a density estimate based on a sample of size $n=10000$ from the blue density, using a too small bandwidth which causes the so-called spurious modes. In this artificial example, simply using a larger bandwidth fixes such a problem, but in a real data scenario it is important to discern which features are really present in the distribution and which are merely artifacts due to sampling fluctuations. In this regard, some authors have developed techniques to test if an apparent mode is significant or not. A representative, though surely incomplete, list of references includes \cite{minnotte1997,fisher2001,dumbgen2008,duong2008,burman2009,genovese2016}.

\begin{figure}\centering
\includegraphics[width=.5\textwidth]{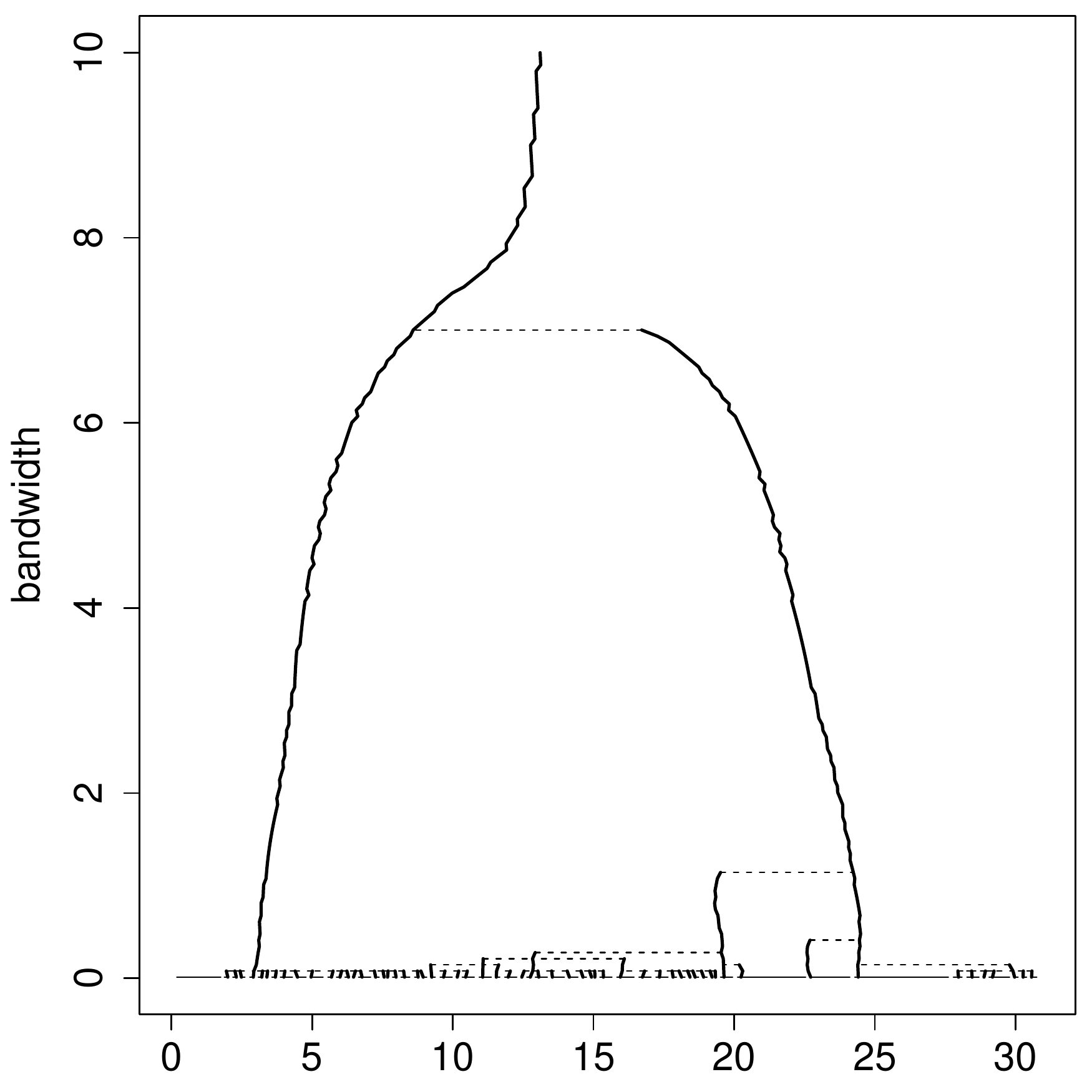}\includegraphics[width=.5\textwidth]{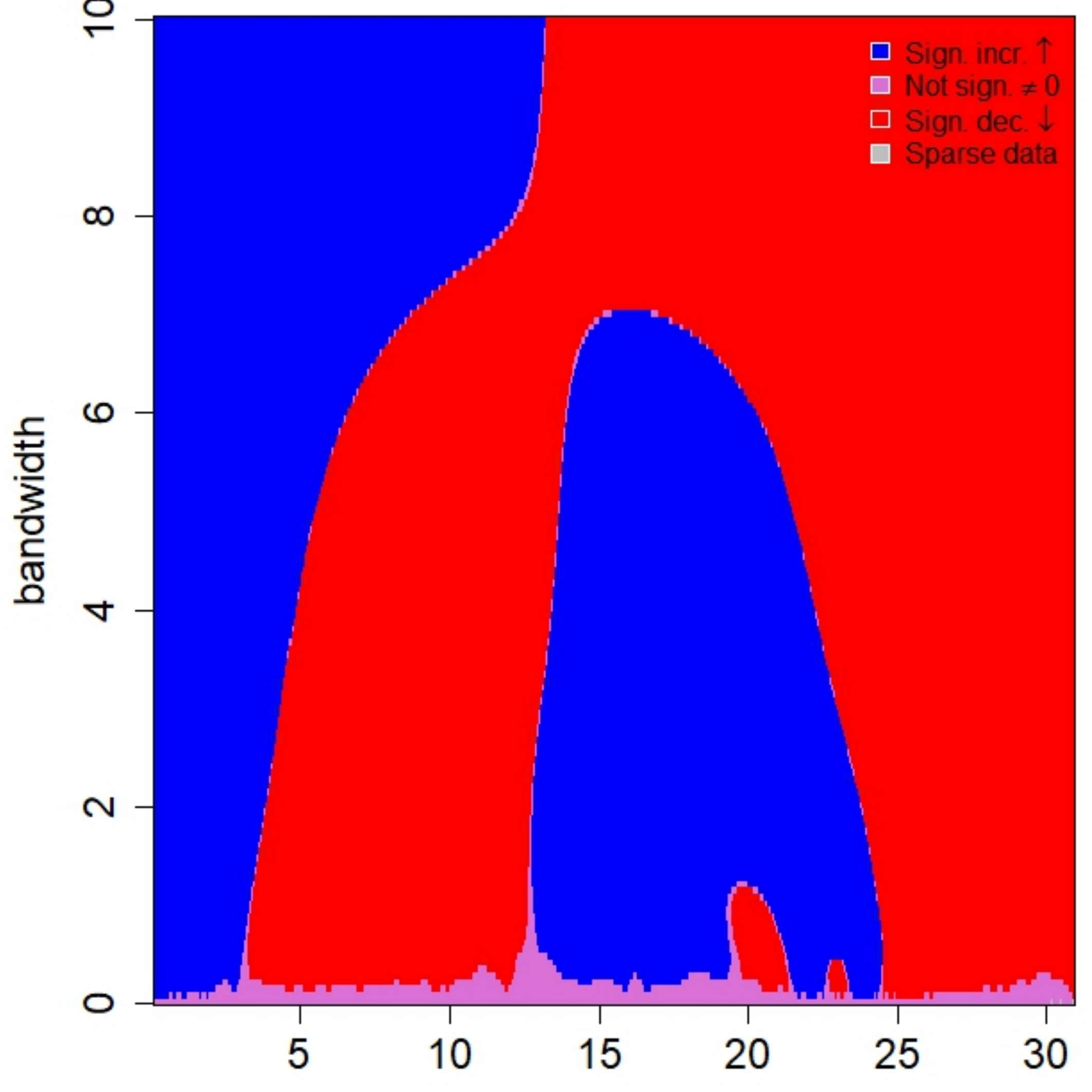}
\caption{Mode tree (left) and SiZer map (right) for the NBA shot data.}
\label{fig:modes}
\end{figure}

Moreover, a number of methodologies are based on investigating how these features evolve as the degree of smoothing varies. To cite a few, let us mention the mode tree \citep{minnotte1993}, the SiZer \citep{chaudhuri1999} and persistence diagrams \citep{cohen2007}. The mode tree depicts the location of the modes of the kernel density estimator as the bandwidth $h$ varies, starting from a large value that produces a unimodal estimate, down to a point where the estimator is severely undersmoothed with numerous small bumps (the number of modes of a normal-kernel density estimator is known to be a non-increasing function of $h$, as shown in \citealp{silverman1981}). Figure \ref{fig:modes} (left) shows the mode tree for the NBA shot data.

SiZer is short for significant zero crossings of the density derivative, which is precisely the event that this tool seeks for. By exploring the locations where such crossings are significant at different levels of resolution (i.e., bandwidths), the SiZer precisely addresses the issue of detecting which density features are ``really there''. Its outcome is a color map on the \{data location\}$\times$\{bandwidth range\} plane (the scale-space approach), with different colors corresponding to regions where the density is significantly increasing (blue), decreasing (red), or where the procedure is inconclusive (purple). The right picture in Figure \ref{fig:modes} shows the SiZer color map for the NBA shot data; it is quite similar to the mode tree, but somehow it displays more information due to the significance analysis. While the two extreme modes are significant for a large range of scales, the existence of the two middle modes is certainly debatable as they look less important. But it should be noted that there is a bandwidth for which all four modes are significant; hence, Figure \ref{fig:NBA} follows the advice of displaying the level of smoothing ``that maximizes the number of significant modes'', as suggested by \cite{genovese2016}.

The persistence diagram, a representation borrowed from the field of Computational Geometry, is one of the fundamental tools in a recently developed branch of Statistics called Topological Data Analysis \citep{wasserman2018}. Each mode has a lifetime, determined by its birth and death. In turn, these are obtained by inspecting the connected components of the upper level sets of the density, $L(c)=\{x\in\mathbb R^d\colon f(x)\geq c\}$ as the level $c$ varies from $\max f(x)$ down to zero. Along this journey, every time the level $c$ reaches the height of a local maximum, a new connected component of $L(c)$ emerges (this is the birth time), and as $c$ decreases some of these connected components eventually merge. When two of them merge, the most recently created component is said to die (so, this is the death time), while the elder one is considered to remain alive.

\begin{figure}\centering
\includegraphics[width=.5\textwidth]{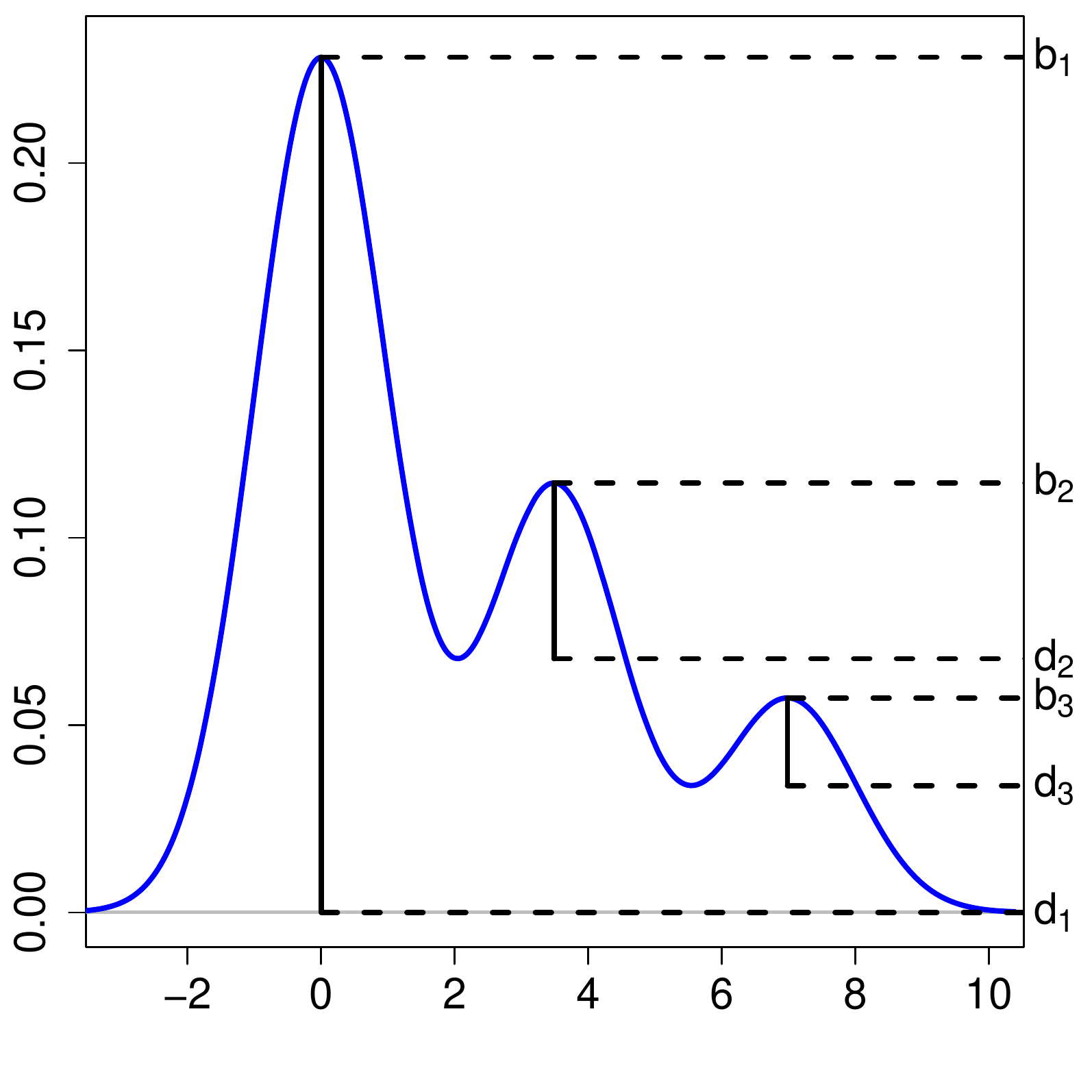}\includegraphics[width=.5\textwidth]{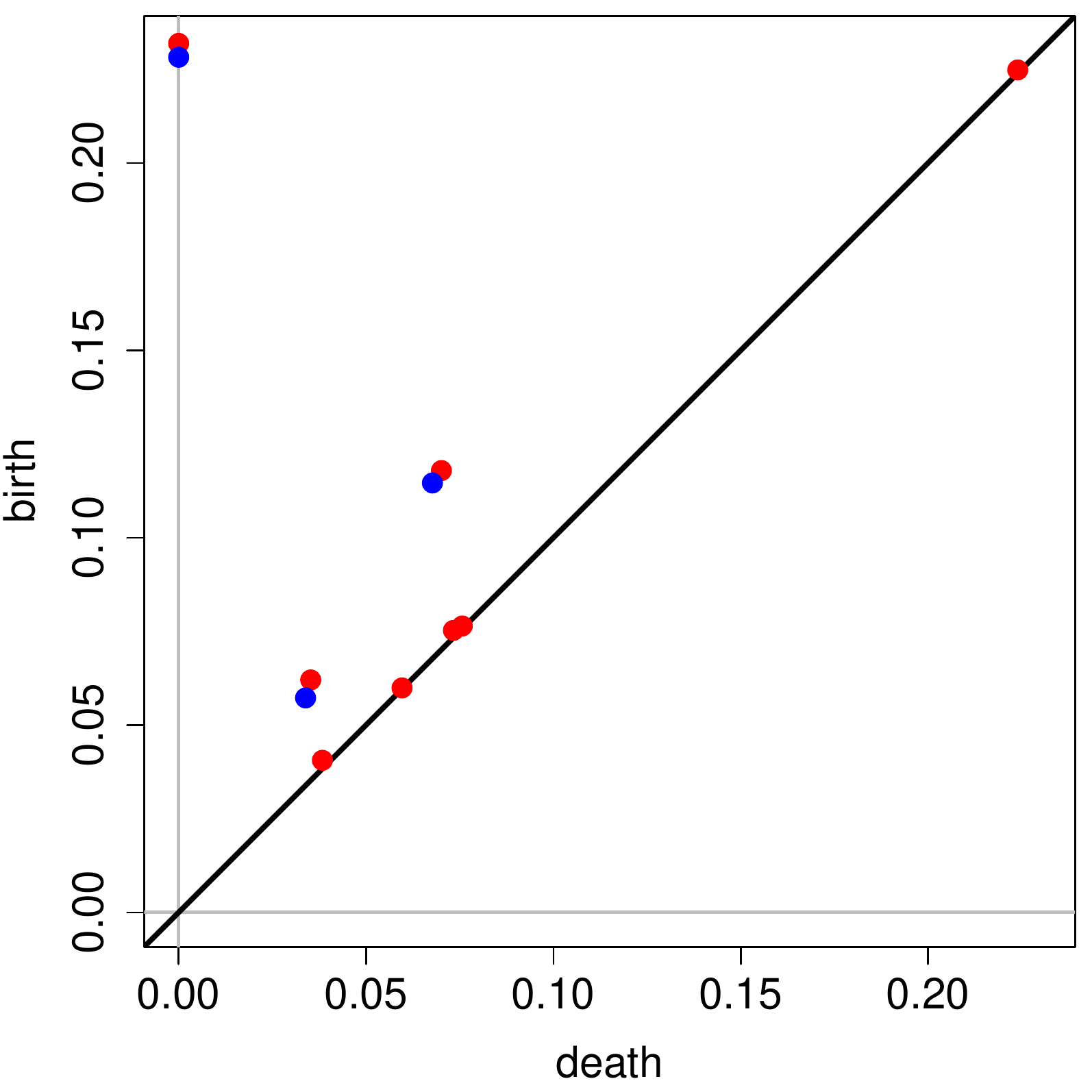}
\caption{Birth-death bars (left) and their representation in a persistence diagram (right).}
\label{fig:persistence}
\end{figure}

The left picture in Figure \ref{fig:persistence} shows the birth and death heights for the blue density in Figure \ref{fig:2dens}. The persistence diagram simply consists of representing the pairs $(d_i,b_i)$ of death and birth heights for each of the density modes. Since $b_i\geq d_i$, all the points lie above the diagonal; moreover, points close to the diagonal correspond to low persistence modes, since their birth and death heights are very close. Hence, a short lifetime is considered as an indication that the corresponding mode is unimportant and may have been caused by the sample variability. The right picture in Figure \ref{fig:persistence} shows three blue points with the death-birth levels of the three modes in the blue density on the left picture, plus eight red dots which correspond to the eight modes in the red density estimate in Figure \ref{fig:2dens}. Despite having a very different number of nodes, the persistence diagram reveals that five out of those eight modes are little persistent, since their death-birth levels stand quite close to the diagonal. In contrast, the remaining three modes appear to be really there, and besides, they have death-birth levels which are very similar to those of the true density modes. For this single-bandwidth case, \cite{fasy2014} showed that it is possible to construct a band above the diagonal to identify the significant modes; on the other hand, \cite{sommerfeld2017} suggested to inspect all the persistence diagrams as the bandwidth varies to evaluate the significance of the modes across different degrees of smoothing.

\section{Modal clustering}\label{sec:modclust}

As noted above, many of the distributions encountered when analyzing real data present hidden structures, which are commonly understood to correspond to the existence of sub-populations. The identification of these sub-groups is precisely the goal of cluster analysis; more formally, it consists of obtaining a meaningful partition $\mathcal C=\{C_1,\dots,C_r\}$ of the whole sample space.

Many different clustering methodologies exist, which mainly differ on the notion of ``meaningful'' in the previous sentence, and a modern compendium of them is given in \cite{hennig2016}. Here we embrace the philosophy of \cite{carlsson2013}, who noted that the ``density needs to be incorporated into the clustering procedures'', and focus on density-based clustering procedures henceforth.

Two main paradigms can be found within density-based clustering approaches, depending on the concept of cluster that it is adopted. The parametric point of view (also called model-based clustering) assumes that the underlying density generating the data is a mixture density $f=\sum_{\ell=1}^L\pi_\ell f_\ell$, as noted in the previous section. Then, any point $x$ in the space can be assigned to the mixture component that makes it more probable by finding the index $\ell\in\{1,\dots,L\}$ that maximizes $\pi_\ell f_\ell(x)$. Therefore, under the mixture model assumption, the ideal clustering is the partition with clusters
\begin{equation}\label{eq:parclus}
C_\ell=\{x\in\mathbb R^d\colon\pi_\ell f_\ell(x)=\max_{j=1,\dots,L}\pi_jf_j(x)\}
\end{equation}
for $\ell=1,\dots,L$.

In contrast, the nonparametric approach makes no assumption about the true density. Instead, clusters are intuitively conceived as regions of concentration of probability mass, separated from each other by lower density zones. However, this natural concept is not so easy to formalize. For a given level $c>0$, \cite{hartigan1975} defined the $c$-clustering as the collection of connected components of the density level set $L(c)=\{x\colon f(x)\geq c\}$. This methodology was extensively analyzed in \cite{CFF00,CFF01} and \cite{rinaldo2010}. Nevertherless, focusing on a single level $c$ may not allow to discover the whole cluster structure of the distribution, so \cite{stuetzle2003} suggested to use the cluster tree instead, that is, the tree that describes the structure of all the $c$-clusterings as the level $c$ varies. \cite{rinaldo2012} provided a deep theoretical analysis of cluster tree estimators.

\begin{figure}\centering
\includegraphics[width=.5\textwidth]{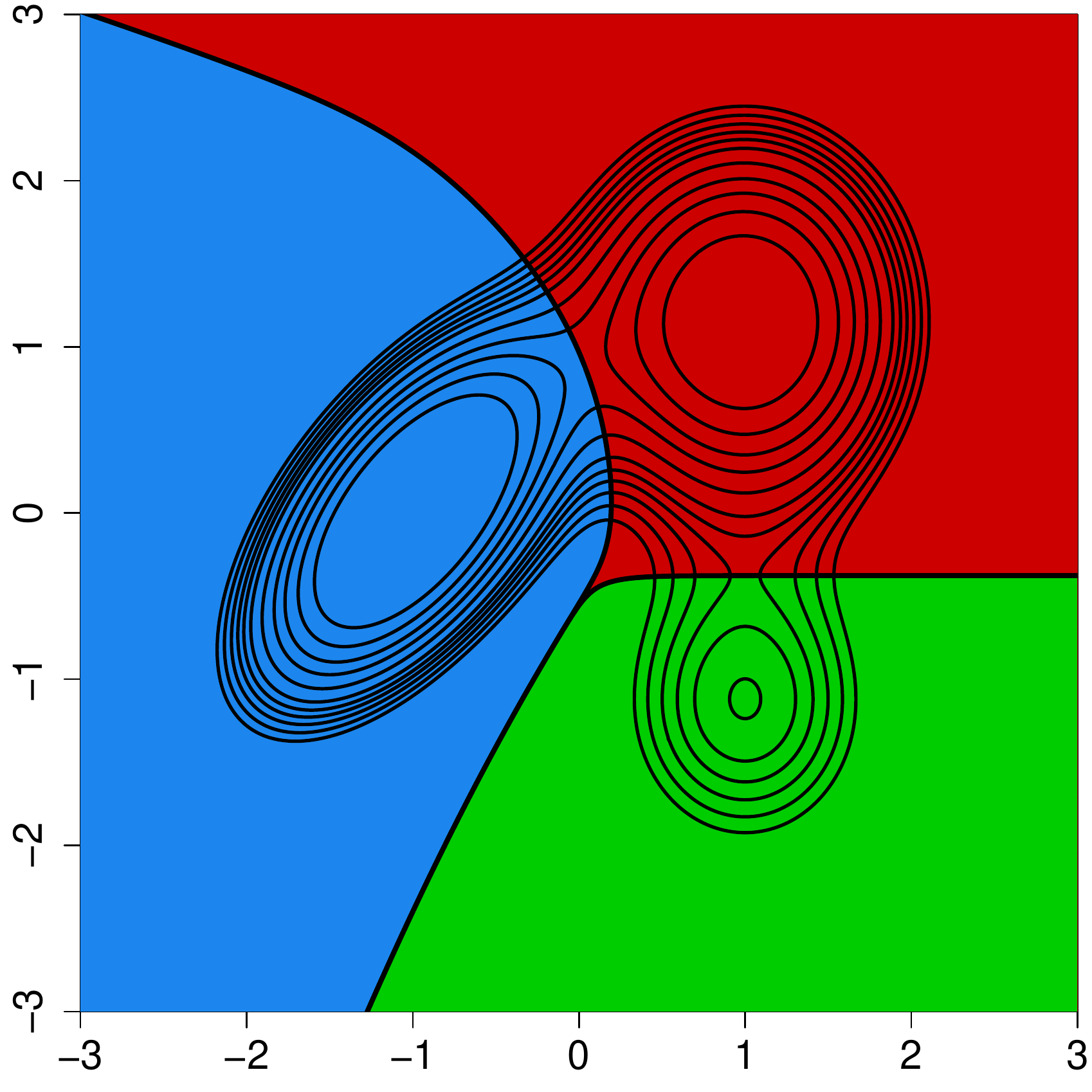}\includegraphics[width=.5\textwidth]{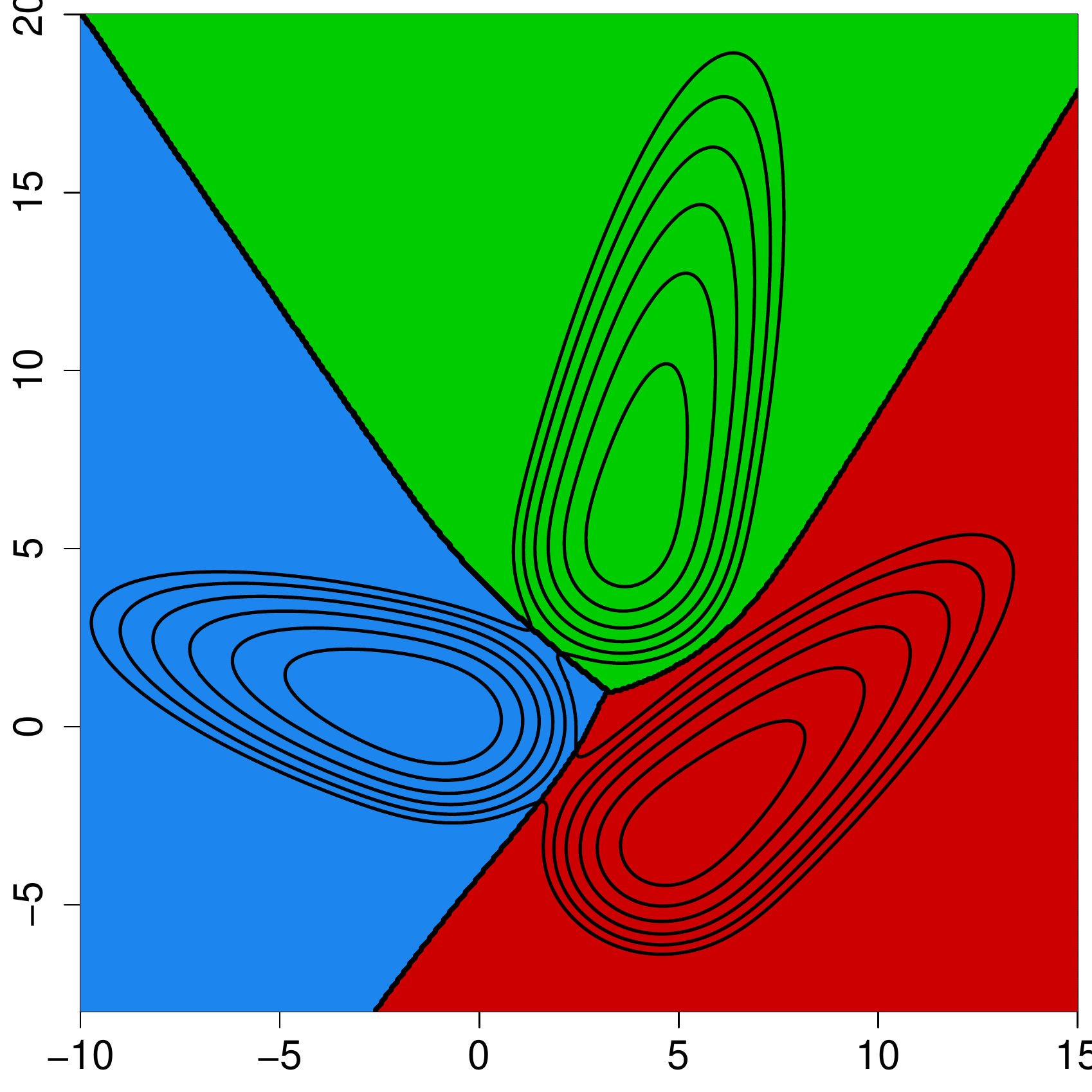}
\caption{Ideal partition of the space in the sense of modal clustering, for two trimodal distributions.}
\label{fig:modclus}
\end{figure}

Alternatively, \cite{chacon2015} showed that this approach can also be formulated in terms of the domains of attraction of the density modes (hence the name of modal clustering): given $x\in\mathbb R^d$, consider the path starting at $x$ that follows the direction of steepest ascent (which is determined by the density gradient $\nabla f$); such a curve $\gamma_x\colon\mathbb R\to\mathbb R^d$ is the solution to the initial value problem $\gamma_x'(t)=\nabla f\big(\gamma_x(t)\big)$ with $\gamma_x(0)=x$. Under certain regularity conditions, that path always finishes at a critical value of $f$, so that $\nabla f\big(\lim_{t\to\infty}\gamma_x(t)\big)=0$ and, therefore, the set of points whose final destination is a given critical point is called the domain of attraction of that critical point. Hence, if $\theta_1,\dots,\theta_r$ denote all the density modes, the ideal clustering from this modal point of view is the partition with clusters
\begin{equation}\label{eq:nonparclus}
C_i=\{x\in\mathbb R^d\colon\lim_{t\to\infty}\gamma_x(t)=\theta_i\}
\end{equation}
for $i=1,\dots,r$. Figure \ref{fig:modclus} shows the density contour plot and the ideal modal clustering (the regions in different colors) for two bivariate trimodal distributions. These examples illustrate that this (admittedly awkward) definition seems to agree with the natural partition dictated by the intuition.

The former two approaches (parametric and nonparametric) are described above at a population level, but of course in practice these ``oracle'' population clusterings need to be estimated from the data. In the model-based setting it is usually assumed that all the component densities belong to a certain parametric family, for instance the multivariate normal, so that given a number of components all the mixture parameters and weights are fitted to the data by maximum likelihood, and the number of components is selected by the Bayesian information criterion \citep[see][]{fraley2002}. This yields estimated clusters by replacing the population quantities by estimates in (\ref{eq:parclus}); here, the number of clusters is the number of estimated mixture components. In modal clustering, the obvious analog consists in approximating the clusters by replacing the true density with a nonparametric density estimator in (\ref{eq:nonparclus}); here, the number of clusters is the number of modes of the density estimate. \cite{chacon2015} showed a set of sufficient conditions on the density estimate to guarantee the consistency of the estimated modal clustering, with respect to a clustering-specific metric.

When the different mixture components are located far enough from each other, the resulting mixture density has as many modes as mixture components \citep{ray2005}. In that case, the clusterings defined by (\ref{eq:parclus}) and (\ref{eq:nonparclus}) are usually very similar. However, the model-based approach gets into trouble when the cluster shapes are markedly different from the density contours of the mixture components; for instance, many normal components may be needed to model a moderately-skewed cluster shape. Then, the identification of clusters and mixture components no longer makes sense, and this has motivated research on the model-based approach to amend this flaw, mainly focusing on component merging techniques to avoid the aforementioned situation \citep[see][and references therein]{hennig2010}. In this regard, modal clustering offers an additional solution: fit a normal mixture model to the data, to obtain an accurate density estimate $\widehat f$, but then combine this mixture-model density estimate with the modal approach (\ref{eq:nonparclus}) to find the clusters according to a modal point of view. Even if several mixture components are used to fit a nonnormal cluster, they will result in a single cluster as long as those components correspond to a unimodal group. This procedure was explored in detail in \cite{chacon2018b}; see also \citet{scrucca2016}.

To compare both modal approaches, consider Wally's data, freely available from Randal Olson's blog\footnote{\tt http://www.randalolson.com/wp-content/uploads/wheres-waldo-locations.csv}. This dataset contains the coordinates of all 68 Wally's locations in the primary seven editions of the popular Martin Handford's \emph{Where's Wally?} puzzle books (Waldo, in the US and Canada edition). The goal here is to look for sub-groups of locations that point out different zones where Wally is usually positioned.

\begin{figure}\centering
\includegraphics[angle=-90,origin=c,height=.5\textwidth]{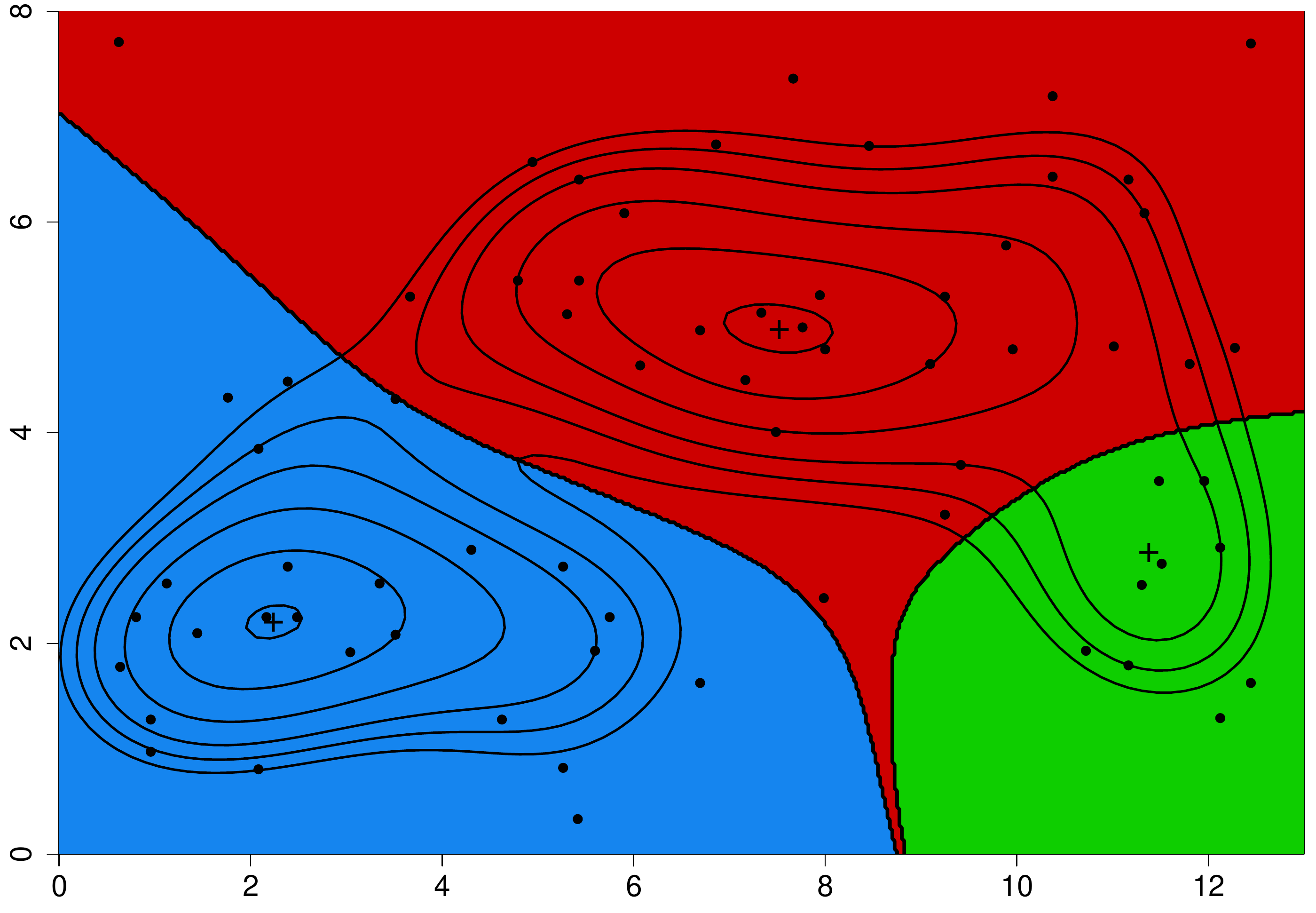}\\[-1cm]
\includegraphics[angle=-90,origin=c,height=.5\textwidth]{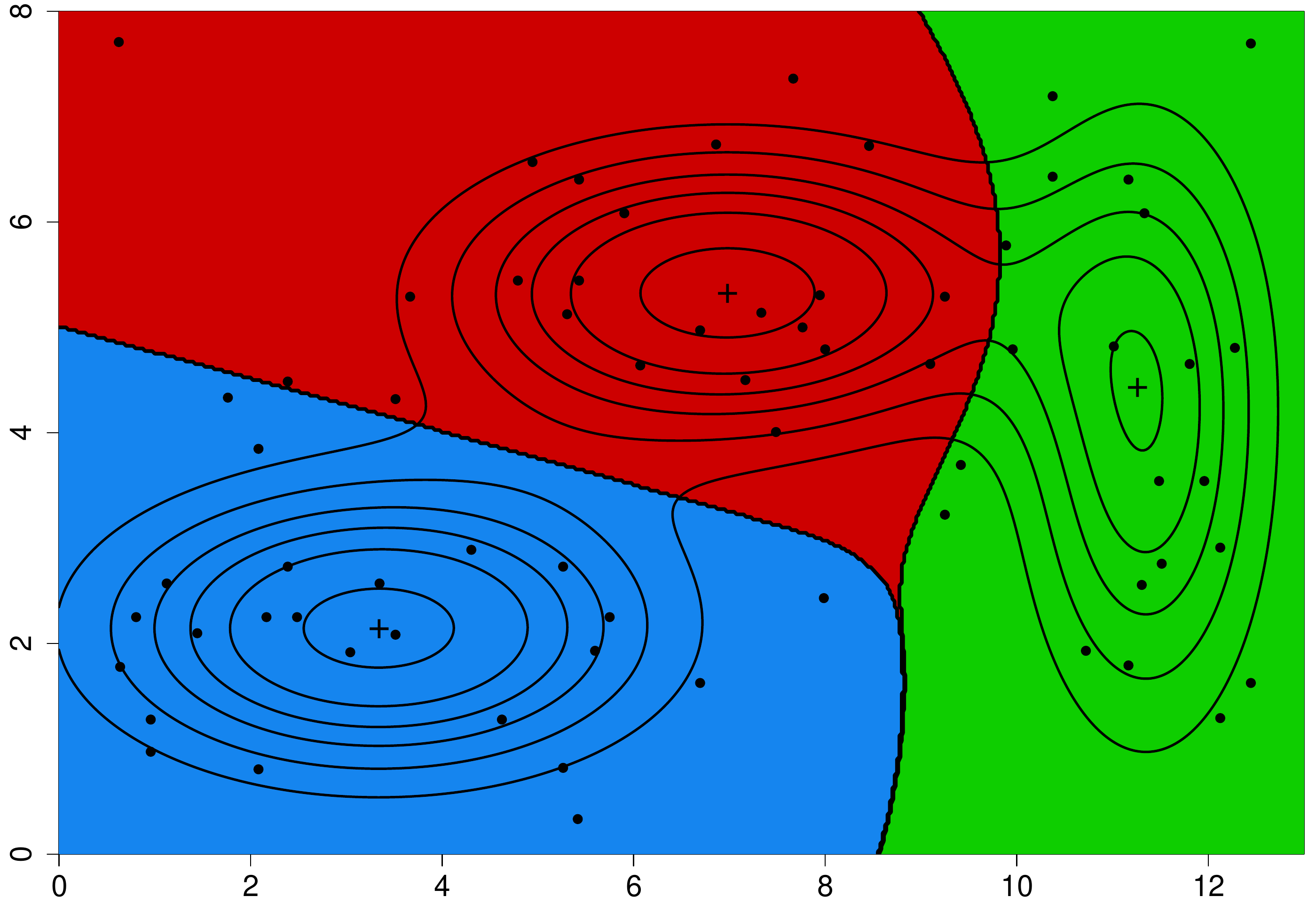}
\caption{Modal clustering for Wally's data, based on a kernel density estimate (top) and based on a normal mixture density estimate (bottom).}
\label{fig:wally}
\end{figure}

Figure \ref{fig:wally} shows the resulting modal clusterings based on a kernel density estimate (top picture) and on a normal mixture density estimate (bottom picture). The dots represent Wally's locations, the contour lines depict the density estimates, the crosses indicate the estimated modes and the colored regions, delimited by black thick lines, constitute the whole-space clusters. Both methodologies agree that there exist three different concentration zones around which Wally is usually placed, since the two density estimates have three modes. The main difference is that, while the component density contours are clearly elliptical for the normal mixture density estimate, the kernel density estimate shows no such restrictions. This implies that they mostly agree on the blue cluster, but present notable differences regarding the partition in the remaining red and green regions.

\section{Modal regression}

All the above sections deal with inference problems about a single (possibly multivariate) random variable $X$. Here, it will be shown that a modal point of view can also be useful when the goal is to study the relationship between a response variable $Y$ and a random vector $X$ of covariates.

Statistically, this relationship is fully described by the conditional distribution of $Y$ given $X=x$ (written $Y|X=x$, for short), and the most common way to summarize this distribution is through the (mean) regression function $m(x)=\mathbb E(Y|X=x)$. However, as learned from the previous sections, the mean is not always the most appropriate summary: if the distribution of $Y|X=x$ is moderately-to-highly skewed, or if it is multimodal, then the conditional mean might not be useful to capture the relationship between $X$ and $Y$. Adopting again a modal point of view, the aim would be to study $\theta(x)$, the (global) mode of $Y|X=x$ or, more generally, $\Theta(x)$, a multivalued function that records all the local maxima of the conditional density. The former was first considered in the pioneering work of \cite{sager1982}, while the latter was introduced in \cite{einbeck2006}. Recently, \cite{chen2018} provided a thorough review of both possibilities and the (relatively scant) existing literature on the topic; see also \cite{chen2016}.

\begin{figure}\centering
\includegraphics[width=.5\textwidth]{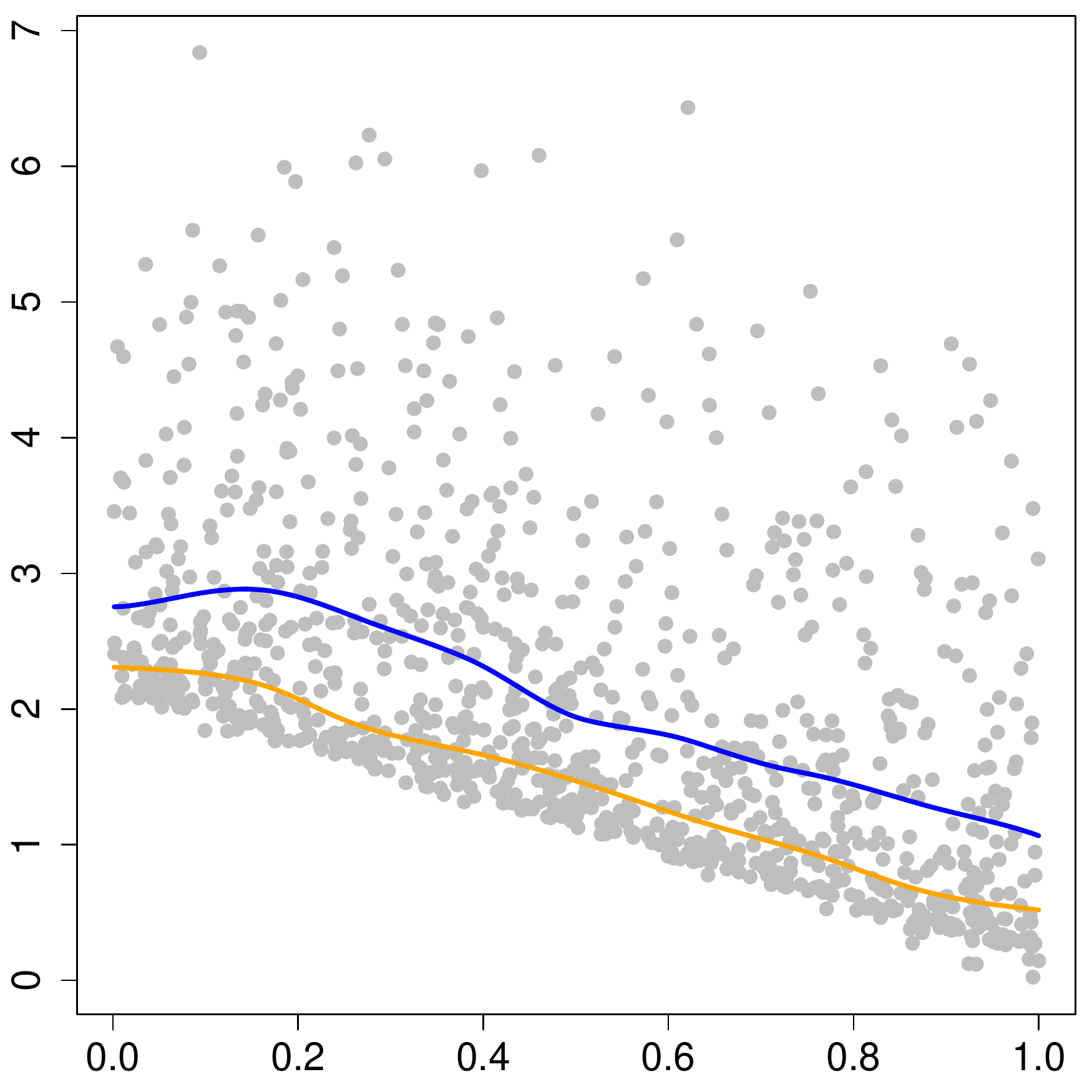}\includegraphics[width=.5\textwidth]{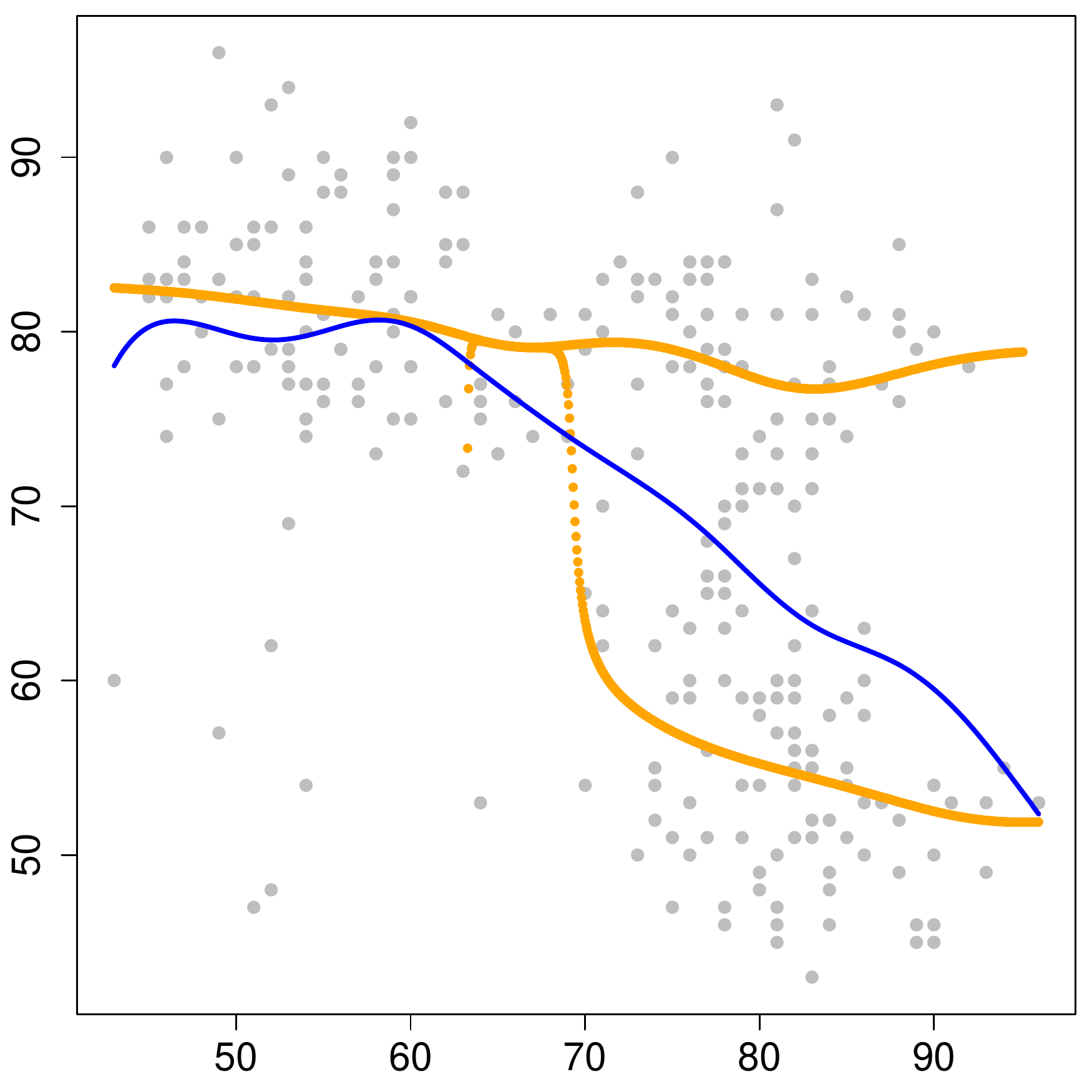}
\caption{Modal regression (orange) versus mean regression (blue).}
\label{fig:modreg}
\end{figure}

Figure \ref{fig:modreg} illustrates these two scenarios. The left picture shows a synthetic sample of size $n=1000$ of a random vector $(X,Y)$ such that $Y|X=x$ is distributed as $5-2x+W$, where $W$ follows the strongly skewed distribution (normal mixture \#3) introduced in \cite{marron1992}. The blue line corresponds to the local linear (mean) regression estimator \citep[see][]{fan1996} which, due to the severe skewness, runs through a relatively uninteresting region regarding the relationship between the two variables. The conditional mode estimate (in orange), however, seems more appropriate here since it points out the most likely value of $Y$ given $X=x$.

The right picture of Figure \ref{fig:modreg} concerns the famous Old Faithful geyser data \citep{azzalini1990}. The two coordinates of each data point represent the waiting time (in minutes) from the previous eruption to the current one and the waiting time from the current eruption to the following one, and the goal is to predict the next waiting time given the previous one. Again, the blue line is the local linear regression estimate and the multivalued modal regression estimate is shown in orange. When the covariate is less than 70, the conditional distribution appears to be unimodal and the two estimates are reasonably close; it is noticeable, though, that the modal approach seems to be more robust, since it is not influenced by the outliers occurring around $X=50$. For $X\geq70$ the modal regression estimate splits into two branches, as it looks like the conditional distribution becomes bimodal, with some of the waiting times there followed by around 55 minutes until the next eruption and some other by approximately 80 minutes. The mean regression line, on the other hand, is unaware of this dichotomy and passes through a relatively low density zone, which would lead to a very unlikely prediction.

\section{Discussion}

A long time has passed by since \cite{pearson1895} coined the term ``mode'' to refer to the most likely value in a distribution. \cite{sager1983} reviewed its role since that first appearance until the early 80s, including an interesting historical perspective and a philosophical debate confronting parametric and nonparametric Statistics. Here, the first steps in mode estimation are revisited, recalling for completeness the admittedly-little-used direct mode estimators proposed by Chernoff, Grenander and Dalenius-Venter. The review is extended up to the current times by elaborating on the more popular indirect estimators, especially those based on a preliminary kernel density estimate.

The exposition focuses on the asymptotic properties of these estimators, but deliberately omits a crucial issue: the data-based selection of the tuning parameters on which these estimators depend (the interval length $a$, the proportion $k/n$ of observations in the interval, the power $p$, the bandwidth $h$, or the number of nearest neighbors). The reason for doing so is that such a problem is much more technical so, since this manuscript was conceived as an expository paper, it was surely preferable to focus on the applications and extensions of this ``modal way of thinking'' to other fields. In addition, there does not exist so many papers that study the performance of mode estimators in practice; \cite{grund1995} and \cite{klemela2005} are notable exceptions that include bandwidth selection proposals for kernel mode estimation and brief simulation studies. As a (very) general recommendation, supported by the close link between the two problems, choosing these parameters based on a global error criterion for density gradient estimation usually yields good practical results for mode estimation as well. Several data-based methodologies for this closely related problem were proposed in \cite{chacon2013}.

Extending the scope of modal analysis starts by considering all density local maxima as modes. The existence of multiple modes is associated with different sub-groups in the data, and raises natural questions, such as: How many are there? Is this a real one? Which is exactly the sub-population associated to this mode? These issues have pushed statisticians to interact with other seemingly unrelated branches of Mathematics, like Computational Geometry, Morse Theory or Algebraic Topology, which have revealed themselves as very useful to develop some modern statistical techniques. For instance, Morse Theory allows to provide a very precise definition for the ideal population goal of modal clustering.

Finally, the modal approach also proves to be highly valuable in problems with covariates, such as the classical regression setting. It agrees with the standard mean regression analysis in symmetric situations, but a modal point of view seems advantageous in skewed scenarios, and resistant to outliers; furthermore, it allows to discover unexpected structures which can not be found by more conventional techniques. And, of course, the benefits of the mode are well known long ago for classification problems: after all, the optimal classifier (the Bayes rule) assigns a new unlabeled observation to the class with the highest posterior probability, given the covariates \citep{devroye1996}. That is, to the modal class of that categorical conditional distribution.

\bigskip

\noindent\textbf{Acknowledgements.} The author is grateful to Rosa Crujeiras for her encouragement to write this article. This work has been partially supported by the Spanish Ministerio de Econom\'{\i}a, Industria y Competitividad grant MTM2016-78751-P.

\bibliographystyle{apalike}
\label{FinArt1}

\end{document}